\documentclass[11pt]{article}
\usepackage[left=1in, right=1in, top=1in, bottom=1.5in]{geometry}

\usepackage{amssymb,latexsym,times,graphicx, amsmath}
\usepackage{hyperref}
\usepackage{xcolor}
\hypersetup{
    colorlinks=true,
    linkcolor=blue,
    filecolor=magenta,      
    urlcolor=teal,
    citecolor=violet
}
\usepackage{tocbibind}

\usepackage{amstext}
\usepackage{amsthm}

\usepackage{tabu}

\usepackage{setspace}
\newtheorem{theorem}{Theorem}[section]
\newtheorem{remark}{Remark}[section]

\newtheorem{proposition}{Proposition}[section]

\newtheorem{corollary}{Corollary}[section]

\numberwithin{equation}{section}

\begin{document}

\title{Well Posed Origin Anywhere Consistent Systems in Celestial Mechanics}
\author{Harry Gingold and Jocelyn Quaintance\\
 {\small West Virginia University, Morgantown, WV, USA; gingold@math.wvu.edu}\\
{\small University of Pennsylvania, Philadelphia, PA, USA; jocelynq@seas.upenn.edu}}

\maketitle
\begin{abstract}
Certain measurements in celestial mechanics necessitate
having the origin $O$ of a Cartesian coordinate system (CCS) coincide
with a point mass. For the two and three body problems we show mathematical
inadequacies in Newton's celestial mechanics equations (NCME) when the origin
of a coordinate system coincides with a point mass. A certain system
of equations of relative differences implied by NCME is free of these
inadequacies and is invariant with respect to any CCS translation.
A new constant of motion is derived for the relative system. It shows
that the universe of relative differences of the $N$-body problem
is ``restless''. 
\end{abstract}


\section{Introduction}
Observations and measurements from earth or from other moving bodies
viewed theoretically as point masses require the utilization of a
Cartesian coordinate system (CCS) in which the origin $O$ coincides with the center of mass of
one of these point masses. Their acceleration is modeled by 
Newton's celestial mechanics equations (NCME), namely
\begin{equation}\label{newtonlawcm}
r''_{i}(t)=\sum_{j\neq i}\frac{Gm_{j}(r_{j}(t)-r_{i}(t))}{\|r_{j}(t)-r_{i}(t)\|^{3}},\;1\leq i\leq N,
\end{equation}
where $r_{i}(t)$ is the position vector (in $\mathbb{R}^{3}$) of
the $i$th body (which has mass $m_{i}$), $\|\,\,\|$ represents
the Euclidean norm, and $G$ is the gravitational constant. Ideally
we want initial value problems of these systems to possess unique
solutions no matter where the origin of the coordinate system is in
space. The following theorem, whose proof is found on Pages 1-7 of \cite{HSIEHSIBUYA}, applies.

\begin{theorem}\label{basicexistence theorem}
Given any CCS, for $1\leq j\leq N$, let $r_{j}(t)\in\mathbb{R}^{3}$
and $r_{j}'(t)=dr_{j}/dt\in\mathbb{R}^{3}$ be the position and velocity
of the jth point mass respectively. Assume that the accelerations
$r_{j}''(t)$ satisfy (\ref{newtonlawcm}). Let $1\leq i,j\leq N$,
with $i\neq j$. Then the initial value problem 
\begin{equation}\label{newtoninitalvaluecm}
r_{j}''(t)=\sum_{{i=1\atop i\neq j}}^{N}\frac{Gm_{i}(r_{i}(t)-r_{j}(t))}{\|r_{i}(t)-r_{j}(t)\|^{3}},
\qquad r_{i}(t_{0}),r_{i}'(t_{0})\in\mathbb{R}^{3},
\quad r_{i}(t_{0})-r_{j}(t_{0})\neq\overrightarrow{0},
\end{equation}
possesses unique solutions $r_{j}(t)\in C^{2}([a,b]),\;a<b$ , on some
interval $[a,b]\subseteq\mathbb{R}$, where $t_{0}\in[a,b]$. 
\end{theorem}

Inconsistencies in NCME occur when the origin $O$ of a CCS coincides
with a point mass. This is discussed in Section 2. We propose in Section 3 to
replace NCME by a related system of equations that is free of inconsistencies.
In Section 4 we derive a new constant of motion for the relative difference equations.
In Section 5 we show by means of  Section 4 that our universe is ``restless''. 
Conclusions in Section 6 provide an overview of our study.

\section{Inconsistencies in Body-Centered Origin Systems (BCOS) for the 2
and 3-Body Problems}
Given a two body problem with two point masses $m_{1}$ and $m_{2}$,
suppose that an origin $O$ of a CCS coincides with the point mass
$m_{1}$. Let the $3\times1$ vectors $r_{1}(t), r_{2}(t)$, or
$r_{1},\,r_{2}$, denote the position vectors of $m_{1}, m_{2}$ respectively.
The NCME are 
\begin{equation}\label{originalNew2}
r_{1}''(t)=\frac{Gm_{2}(r_{2}-r_{1})}{\|r_{2}-r_{1}\|^{3}},\qquad r''_{2}(t)
=\frac{Gm_{1}(r_{1}-r_{2})}{\|r_{1}-r_{2}\|^{3}},
\end{equation}
where $\|r(t)\|=\sqrt{r^{T}r}$. Measurements made from the point
$O$ require that we set in (\ref{originalNew2}) 
\begin{equation}\label{r1origin}
r_{1}(t)\equiv\overrightarrow{0}\Longrightarrow r'_{1}(t)\equiv r''_{1}(t)\equiv\overrightarrow{0},
\end{equation}
where $\overrightarrow{0}=[0,0,0]^{T}$. The second equation of (\ref{originalNew2})
becomes 
\begin{equation}
r''_{2}(t)=-\frac{Gm_{1}r_{2}}{\|r_{2}\|^{3}}.\label{Keplerprob}
\end{equation}
But what happens to the first equation of (\ref{originalNew2})? It
implies
\begin{equation}\label{r1diffeq}
\overrightarrow{0}=\frac{Gm_{2}r_{2}}{\|r_{2}\|^{3}}\Longrightarrow0
=\left\Vert \overrightarrow{0}\right\Vert =\left\Vert \frac{Gm_{2}r_{2}}{\|r_{2}\|^{3}}\right\Vert 
=\frac{Gm_{2}}{\|r_{2}\|^{2}}\Longleftrightarrow\|r_{2}\|=\infty.
\end{equation}
Evidently, this is a contradiction to a well known solution $r_{2}(\theta)=\frac{ed}{1+e\cos\theta}$
to (\ref{Keplerprob}). One may explain the inconstancy by claiming
that the origin O is accelerating and try to introduce fictitious
forces to avoid this. In contrast one may blame the inconsistency
on overdetermination inherent in BCOS. Therefore we subtract the two
equations in (\ref{originalNew2}) and propose one origin independent
relative system 
\begin{equation}\label{N2relativesys}
(r_{1}-r_{2})''=-\frac{G(m_{1}+m_{2})(r_{1}-r_{2})}{\|r_{1}-r_{2}\|^{3}}.
\end{equation}
Thus, when $r_{1}(t)\equiv\overrightarrow{0}$, i.e. when $m_{1}$ corresponds
to origin $O$, Equation (\ref{N2relativesys}) becomes a modified
Kepler problem
\begin{equation}\label{modKeplerprob}
-r''_{2}(t)=\frac{G(m_{1}+m_{2})r_{2}}{\|r_{2}\|^{3}}.
\end{equation}

Now we analyze a BCOS model for the $3$-body problem. From a fixed
origin $O$ we record the position of three point masses, $m_{1},m_{2}$,
and $m_{3}$ as the $3\times1$ vectors $r_{1}(t)$, $r_{2}(t)$,
and $r_{3}(t)$ respectively. Then NCME give rise to the following
system of three nonlinear second order differential equations:
\begin{align*}
r''_{1}(t) & =\frac{Gm_{2}(r_{2}-r_{1})}{\|r_{2}-r_{1}\|^{3}}+\frac{Gm_{3}(r_{3}-r_{1})}{\|r_{3}-r_{1}\|^{3}}\\
r''_{2}(t) & =\frac{Gm_{1}(r_{1}-r_{2})}{\|r_{1}-r_{2}\|^{3}}+\frac{Gm_{3}(r_{3}-r_{2})}{\|r_{3}-r_{2}\|^{3}}\\
r''_{3}(t) & =\frac{Gm_{1}(r_{1}-r_{3})}{\|r_{1}-r_{3}\|^{3}}+\frac{Gm_{2}(r_{2}-r_{3})}{\|r_{2}-r_{3}\|^{3}}.
\end{align*}
Assume $m_{1}$ corresponds to $O$, i.e. $r_{1}\equiv\overrightarrow{0}$.
Since $r_{1}\equiv\overrightarrow{0}$ implies that $r'_{1}\equiv r''_{1}\equiv\overrightarrow{0}$,
the preceding system becomes
\begin{align*}
\overrightarrow{0} & =\frac{Gm_{2}r_{2}}{\|r_{2}\|^{3}}+\frac{Gm_{3}r_{3}}{\|r_{3}\|^{3}}\\
r''_{2}(t) & =-\frac{Gm_{1}r_{2}}{\|r_{2}\|^{3}}+\frac{Gm_{3}(r_{3}-r_{2})}{\|r_{3}-r_{2}\|^{3}}\\
r''_{3}(t) & =-\frac{Gm_{1}r_{3}}{\|r_{3}\|^{3}}+\frac{Gm_{2}(r_{2}-r_{3})}{\|r_{2}-r_{3}\|^{3}}.\quad 
\tag{$S1$}
\end{align*}
The first equation of $(S1)$ implies that 
\begin{equation}\label{r2r3first}
\frac{m_{2}r_{2}}{\|r_{2}\|^{3}}
=-\frac{m_{3}r_{3}}{\|r_{3}\|^{3}}\Longleftrightarrow r_{2}=-\frac{m_{3}\|r_{2}\|^{3}}{m_{2}\|r_{3}\|^{3}}r_{3}.
\end{equation}
If we take the norm of (\ref{r2r3first}) we find that 
\begin{equation}
\|r_{2}\|^{-2}=\frac{m_{3}}{m_{2}}\|r_{3}\|^{-2}\Longleftrightarrow\|r_{2}\|=\sqrt{\frac{m_{2}}{m_{3}}}\|r_{3}\|.\label{r2r3norm}
\end{equation}
By substituting (\ref{r2r3norm}) into (\ref{r2r3first}) we obtain
the relation 
\begin{equation}\label{r2r3second}
r_{2}=-\sqrt{\frac{m_{2}}{m_{3}}}r_{3}.
\end{equation}
We now place (\ref{r2r3first}) and (\ref{r2r3second}) into the second equation of $(S1)$
to obtain 
\begin{equation}\label{r3accelfirst}
r''_{3}=-G\left(\frac{m_{3}}{m_{2}}\right)^{\frac{3}{2}}
\left[\frac{m_{1}(\sqrt{m_{3}}+\sqrt{m_{2}})^{2}+m_{2}m_{3}}{(\sqrt{m_{3}}+\sqrt{m_{2}})^{2}}\right]\frac{r_{3}}{\|r_{3}\|^{3}}.
\end{equation}
However, if we substitute (\ref{r2r3second}) into the third equation of $(S1)$ we obtain
\begin{equation}\label{r3accelsec}
r''_{3}=-G\left[\frac{m_{1}(\sqrt{m_{3}}+\sqrt{m_{2}})^{2}+m_{2}m_{3}}{(\sqrt{m_{3}}+\sqrt{m_{2}})^{2}}\right]\frac{r_{3}}{\|r_{3}\|^{3}}.
\end{equation}
Since (\ref{r3accelfirst}) must equal (\ref{r3accelsec}) we have
the relation \footnotesize{
\begin{equation}\label{r3accelrel}
\left(\frac{m_{3}}{m_{2}}\right)^{\frac{3}{2}}
\left[\frac{m_{1}(\sqrt{m_{3}}+\sqrt{m_{2}})^{2}+m_{2}m_{3}}{(\sqrt{m_{3}}+\sqrt{m_{2}})^{2}}\right]\frac{r_{3}}{\|r_{3}\|^{3}}
=\left[\frac{m_{1}(\sqrt{m_{3}}+\sqrt{m_{2}})^{2}+m_{2}m_{3}}{(\sqrt{m_{3}}+\sqrt{m_{2}})^{2}}\right]\frac{r_{3}}{\|r_{3}\|^{3}}.
\end{equation}} 
\normalsize{In} order for (\ref{r3accelrel}) to be valid, either one of two
possibilities occurs. First $\frac{r_{3}}{\|r_{3}\|^{3}}\equiv\overrightarrow{0}$,
which implies that $\|r_{3}\|=\infty$, a contradiction to the assumption
of finite initial conditions, or 
\begin{equation}
\left(\frac{m_{3}}{m_{2}}\right)^{\frac{3}{2}}=1\Longleftrightarrow m_{3}=m_{2}.\label{m2m3equal}
\end{equation}
Thus if $(S1)$ does not satisfy (\ref{m2m3equal}),
NCEM are inconsistent 
with respect to a BCOS model. However $(S1)$ does
satisfy (\ref{m2m3equal}), we can substitute (\ref{m2m3equal}) into
(\ref{r2r3second}) and (\ref{r2r3norm}) to obtain 
\begin{equation}\label{r2r3antipode}
r_{2}(t)=-r_{3}(t)\qquad\text{and}\qquad\|r_{2}(t)\|=\|r_{3}(t)\|.
\end{equation}
Then $(S1)$ reduces to 
\begin{equation}\label{singler3eq}
r_{1}\equiv\overrightarrow{0}\qquad\text{and}\qquad r''_{2}
=-r''_{3}=G\left[m_{1}+\frac{m_{3}}{4}\right]\frac{r_{3}}{\|r_{3}\|^{3}},
\end{equation}
and the second equation of (\ref{singler3eq}) is a modified Kepler
equation which has a conic section curve as a solution. Furthermore,
if $r_{1}\equiv\overrightarrow{0}$ and $r_{2}$ and $r_{3}$ satisfy
(\ref{r2r3antipode}), we see that the center of mass of the three
bodies coincides with $m_{1}$.
\[
\frac{m_{1}r_{1}+m_{2}r_{2}+m_{3}r_{3}}{m_{1}+m_{2}+m_{3}}=\overrightarrow{0}=r_{1}.
\]
The above discussion shows that we can not freely choose the origin
$O$ to be centered on a point mass $m_{1}$, since unless the other
point masses obey ``antipodal" symmetry conditions, the choice of
$r_{1}\equiv\overrightarrow{0}$ in $(S1)$
leads to the conclusion of $m_{2}$ and $m_{3}$ escaping to ``infinity".
To avoid this ``seeming'' inconsistency, we propose to use an origin invariant model of relative
differences, namely 
\begin{align*}
(r_{1}-r_{2})'' & =-\frac{G(m_{1}+m_{2})(r_{1}-r_{2})}{\|r_{1}-r_{2}\|^{3}}-\frac{Gm_{3}(r_{1}-r_{3})}{\|r_{1}-r_{3}\|^{3}}
+\frac{Gm_{3}(r_{2}-r_{3})}{\|r_{2}-r_{3}\|^{3}}\\
(r_{1}-r_{3})'' & =-\frac{Gm_{2}(r_{1}-r_{2})}{\|r_{1}-r_{2}\|^{3}}-\frac{G(m_{1}
+m_{3})(r_{1}-r_{3})}{\|r_{1}-r_{3}\|^{3}}-\frac{Gm_{2}(r_{2}-r_{3})}{\|r_{2}-r_{3}\|^{3}}.
\end{align*}
By setting in $r_{1}(t)\equiv r_{1}'(t)\equiv r_{1}''(t)\equiv\overrightarrow{0}$,
we obtain
\begin{align*}
r''_{2} & =-\frac{G(m_{1}+m_{2})r_{2}}{\|r_{2}\|^{3}}-\frac{Gm_{3}r_{3}}{\|r_{3}\|^{3}}-\frac{Gm_{3}(r_{2}-r_{3})}{\|r_{2}-r_{3}\|^{3}}\\
-r''_{3} & =\frac{Gm_{2}r_{2}}{\|r_{2}\|^{3}}+\frac{G(m_{1}+m_{3})r_{3}}{\|r_{1}-r_{3}\|^{3}}-\frac{Gm_{2}(r_{2}-r_{3})}{\|r_{2}-r_{3}\|^{3}}.
\end{align*}
If $m_{2}$ and $m_{3}$ are small compared to $m_{1}$, the preceding equations
 are perturbations of the second and third equations of $(S1)$. 

\section{Well Posed Origin Anywhere Consistent Systems (WPOACS)}
As the examples of the previous section demonstrate, it is desirable
to have a system of differential equations for the N-body problem
that has the following properties: 
\begin{itemize}
\item{i)} The system of differential equations
is consistent with a CCS with any origin $O$.
\item{ii)} A unique solution is guaranteed. 
\end{itemize}
We call and denote such systems as WPOACS. We
now generalize the origin invariant model of the
relative differences to $N$ point masses $m_{i}$, where $1\leq i\leq N$.
\begin{theorem}\label{origininvariantNbodythm}
Let $O$ be any origin and $r_{i}(t)\in\mathbb{R}^3$ be the position of $m_{i}$.
Denote $N-1$ dependent variables by 
\begin{equation}\label{r_ikdef}
r_{1k}(t):=r_{1}(t)-r_{k}(t),\qquad 2\leq k \leq N.
\end{equation}
Then the following system
\begin{align*}
(r_{1}-r_{k})'' & =G\sum_{i=2}^{N}\frac{m_{i}(r_{i}-r_{1})}{\|r_{i}-r_{1}\|^{3}}
-G\sum_{{i=1\atop i\neq k}}^{N}\frac{m_{i}(r_{i}-r_{k})}{\|r_{i}-r_{k}\|^{3}}\\
 & \hspace{-0.6in}=G\sum_{i=2}^{N}\frac{m_{i}(r_{i}-r_{1})}{\|r_{i}-r_{1}\|^{3}}
 -G\sum_{{i=1\atop i\neq k}}^{N}\frac{m_{i}[r_{1}-r_{k}-(r_{1}-r_{i})]}{\|r_{1}-r_{k}-(r_{1}-r_{i})\|^{3}},\qquad2\leq k\leq N,
 \quad\tag{$RS1$}
\end{align*}
 is invariant with respect to an arbitrary
CCS translation and is WPOACS. 
\end{theorem}
{\bf Proof:} For any $c(t)\in C^2(\mathbb{R})$, observe that
\begin{align*}
r''_{1}-r''_{k} & =(r_{1}+c(t))''-(r_{k}+c(t))''\\
 & \hspace{-0.6in}=\sum_{i=2}^{N}\frac{Gm_{i}(r_{i}+c(t)-(r_{1}+c(t)))}{\|r_{i}+c(t)-(r_{1}+c(t))\|^{3}}
 -\sum_{{i=1\atop i\neq k}}^{N}\frac{Gm_{i}(r_{i}+c(t)-(r_{k}+c(t)))}{\|r_{i}(t)+c(t)-(r_{k}+c(t))\|^{3}}\\
 & =G\sum_{i=2}^{N}\frac{m_{i}(r_{i}-r_{1})}{\|r_{i}-r_{1}\|^{3}}
 -G\sum_{{i=1\atop i\neq k}}^{N}\frac{m_{i}(r_{i}-r_{k})}{\|r_{i}-r_{k}\|^{3}}.
\end{align*}
Thus our system is invariant under an arbitrary translation. Next
we impose on $(RS1)$ the following initial conditions $r_{1k}(t_{0}),\,r_{1k}'(t_{0})\in\mathbb{R}^3$
subject to 
\begin{equation}\label{eq:INITIALCONDITRELATIVEDIFFERENCES}
r_{1k}(t_{0})\neq\overrightarrow{0},\qquad
r_{1k}(t_{0})-r_{1j}(t_{0})\neq\overrightarrow{0},\qquad 1\leq k,j\leq N,\:k\neq j.
\end{equation}
By Pages 1-7 of \cite{HSIEHSIBUYA} there
exists an interval $[a,b]\subseteq\mathbb{R}$, where $t_{0}\in[a,b]$,
on which a solution $r_{1k}(t)\in C^{2}([a,b])$ to $(RS1)$
exists. \qquad $\Box$

\medskip
Observe the perfect match of certain constraints on the initial conditions
in (\ref{newtonlawcm}) to certain constraints on the initial conditions
in (\ref{eq:INITIALCONDITRELATIVEDIFFERENCES}).
\begin{equation}\label{eq:MathICNCME=000026ICRELATICE}
r_{k}(t_{0})-r_{j}(t_{0})\neq\overrightarrow{0}\,\Longleftrightarrow
r_{1k}(t_{0})\neq\overrightarrow{0},\,\,\text{and}
\,\,\:r_{1k}(t_{0})-r_{1j}(t_{0})\neq\overrightarrow{0},\,\,\text{when}\,\,k\neq j,
\end{equation}
where we implicitly assumed that $ 2\leq k, j\leq N$.  Moreover,
the substitution $r_{1}(t)\equiv r_{1}'(t)\equiv r_{1}''(t)\equiv\overrightarrow{0}$
in $(RS1)$ results in a system of equations free of inconsistencies
\begin{equation}\label{reducediffsys}
-r''_{k}=G\sum_{i=2}^{N}\frac{m_{i}r_{i}}{\|r_{i}\|^{3}}+\frac{Gm_{1}r_{k}}{\|r_{k}\|^{3}}
-G\sum_{{i=2\atop i\neq k}}^{N}\frac{m_{i}(r_{i}-r_{k})}{\|r_{i}-r_{k}\|^{3}},\qquad2\leq k\leq N.
\end{equation}

Another system closely related to $(RS1)$ and
invariant under arbitrary translations $c(t)$ is 
\[
(r_{j}-r_{k})''=G\sum_{{i=1\atop i\neq j}}^{N}\frac{m_{i}(r_{i}-r_{j})}{\|r_{i}-r_{j}\|^{3}}
-G\sum_{{i=1\atop i\neq k}}^{N}\frac{m_{i}(r_{i}-r_{k})}{\|r_{i}-r_{k}\|^{3}},\qquad1\leq j<k\leq N,
\tag{$RS2$}
\]
where the dependent variables are 
\begin{equation}\label{rjkdefA}
r_{jk}:=r_{j}(t)-r_{k}(t),\qquad 1\leq j<k\leq N.
\end{equation}
System ($RS2$) consists of $\binom{N}{2}$
equations and is generated by the $N-1$ equations $(r_{1}-r_{p})'',2\leq p\leq N$
since 
\begin{equation}
(r_{j}-r_{k})''=(r_{1}-r_{k})''-(r_{1}-r_{j})''\qquad\text{whenever}\,\,j\neq1.\label{subtractivestep}
\end{equation}
Because the right side of the equations of (\ref{newtonlawcm}) involves
differences of the form $r_{i}-r_{j}$, $(RS2)$
could be technically preferable to $(RS1)$.

\section{A New Origin Invariant Constant of Motion}
In this section we derive a new constant of motion
for $(RS2)$ which has several ramifications that we later discuss..
\begin{theorem}\label{Inequalitythm}
Assume that $r_{j}(t)-r_{k}(t)\in\mathbb{R}^3$, where $1\leq j<k\leq N$, are solutions
on $(a,b)$ to the system of equations 
\begin{align}\label{newtonseconddiff}
(r_{j}-r_{k})'' & =G\sum_{{i=1\atop i\neq j}}^{N}\frac{m_{i}(r_{i}-r_{j})}{\|r_{i}-r_{j}\|^{3}}
-G\sum_{{i=1\atop i\neq k}}^{N}\frac{m_{i}(r_{i}-r_{k})}{\|r_{i}-r_{k}\|^{3}}\nonumber \\
 & =\frac{-G(m_{j}+m_{k})(r_{j}-r_{k})}{\|r_{j}-r_{k}\|^{3}}
 +G\sum_{{i=1\atop i\neq j,k}}^{N}m_{i}\left[\frac{r_{i}-r_{j}}{\|r_{i}-r_{j}\|^{3}}-\frac{r_{i}-r_{k}}{\|r_{i}-r_{k}\|^{3}}\right].
\end{align}
Let $M:= \sum_{i=1}^Nm_i$.
Then the following identity holds
\begin{equation}\label{eq:NoConstantSolRelative}
\sum_{1\leq j<k\leq N}m_{j}m_{k}(r_{j}-r_{k})\cdot(r_{j}-r_{k})''
=-GM\sum_{1\leq j<k\leq N}\frac{m_{j}m_{k}}{\|r_{j}-r_{k}\|}<0.
\end{equation}
\end{theorem}
{\bf Proof:}
Take each equation in (\ref{newtonseconddiff}) and form the
dot product with the vector $m_{j}m_{k}(r_{j}-r_{k})$ to obtain 
\begin{equation}\label{dotproducteq}
m_{j}m_{k}(r_{j}-r_{k})\cdot(r_{j}-r_{k})''=-\frac{Gm_{j}m_{k}(m_{j}+m_{k})}{\|r_{j}-r_{k}\|}+T(j,k),
\end{equation}
where 
\begin{equation}\label{Tjkdef}
T(j,k):=Gm_{j}m_{k}\sum_{{i=1\atop i\neq j,k}}^{N}m_{i}\left[\frac{(r_{j}-r_{k})\cdot(r_{i}-r_{j})}{\|r_{i}-r_{j}\|^{3}}
-\frac{(r_{j}-r_{k})\cdot(r_{i}-r_{k})}{\|r_{i}-r_{k}\|^{3}}\right].
\end{equation}
Next sum both sides of (\ref{dotproducteq}) with
respect to all pairs of indices $(j,k)$ to obtain 
\small{
\begin{equation}\label{diffformkineticengy}
\sum_{1\leq j<k\leq N}m_{j}m_{k}(r_{j}-r_{k})\cdot(r_{j}-r_{k})''
=-\sum_{1\leq j<k\leq N}\frac{Gm_{j}m_{k}(m_{j}+m_{k})}{\|r_{j}-r_{k}\|}+\sum_{1\leq j<k\leq N}T(j,k),
\end{equation}}
\normalsize{where}
\small{\begin{align}\label{expandedT2sum}
 & \sum_{1\leq j<k\leq N}T(j,k)=G\sum_{1\leq j<k\leq N}m_{j}m_{k}(r_{j}-r_{k})\cdot
 \sum_{{i=1\atop i\neq j,k}}^{N}m_{i}\left[\frac{r_{i}-r_{j}}{\|r_{i}-r_{j}\|^{3}}-\frac{r_{i}-r_{k}}{\|r_{i}-r_{k}\|^{3}}\right].
\end{align}}
\normalsize{The} goal is to show that 
\begin{equation}\label{simplifedT2jksum}
\sum_{1\leq j<k\leq N}T(j,k)=-G\sum_{1\leq j<k\leq N}\frac{(M-m_{j}-m_{k})m_{j}m_{k}}{\|r_{j}-r_{k}\|}.
\end{equation}
In order to prove (\ref{simplifedT2jksum}), temporarily fix an index
pair $(j,k)$ and recall that we are summing the $\binom{N}{2}$ equations
$m_{\ell}m_{p}(r_{\ell}-r_{p})\cdot(r_{\ell}-r_{p})''$, where $1\leq\ell<p\leq N$.
Look at the sum of the $\binom{N}{2}-1$ equations arising from $(r_{\ell}-r_{p})''$,
where $1\leq\ell<p\leq N$ and $(j,k)\neq(\ell,p)$, and add together
the terms which have a factor of $\frac{r_{j}-r_{k}}{\|r_{j}-r_{k}\|^{3}}$.
This process is equivalent to interchanging the order of summation
on the right side of (\ref{expandedT2sum}). Observe that in order
to obtain $\frac{r_{j}-r_{k}}{\|r_{j}-r_{k}\|^{3}}$ either $\ell=j,k$
or $p=j,k$. The number of such ordered pairs $(\ell,p)\neq(j,k)$
for which either $\ell=j,k$ or $p=j,k$ is 
$\binom{N}{2}-\binom{N-2}{2}-1=2N-4$,
where $\binom{N-2}{2}$ counts those ordered pairs which are independent
of $j$ and $k$. These $2N-4$ ordered pairs will be paired up
with opposite signs. To explain how the pairing occurs we analyze
four mutually exclusive cases.

\medskip
Case 1: $\ell=j$ and $p\neq k,j$. This implies that $p>j$.
Then $(r_{\ell}-r_{p})''=(r_{j}-r_{p})''$ where
\begin{equation}\label{eqjp}
(r_{j}-r_{p})''=\frac{-G(m_{j}+m_{p})(r_{j}-r_{p})}{\left\Vert r_{j}-r_{p}\right\Vert ^{3}}
+G\sum_{{i=1\atop i\neq j,p}}^{N}\frac{m_{i}(r_{i}-r_{j})}{\left\Vert r_{i}-r_{j}\right\Vert ^{3}}
-G\sum_{{i=1\atop i\neq j,p}}^{N}\frac{m_{i}(r_{i}-r_{p})}{\left\Vert r_{i}-r_{p}\right\Vert ^{3}}.
\end{equation}
Since $p\neq j,k$, only $i=k$ in second
summand on the right side of (\ref{eqjp}) gives rise to $r_{j}-r_{k}$,
in which case we obtain the term
\[
\frac{Gm_{k}(r_{k}-r_{j})}{\left\Vert r_{k}-r_{j}\right\Vert ^{3}}
=-\frac{Gm_{k}(r_{j}-r_{k})}{\left\Vert r_{j}-r_{k}\right\Vert ^{3}}.
\]
We then form the dot product with the vector $m_{j}m_{p}(r_{j}-r_{p})$
to obtain a summand of the form 
\begin{equation}\label{jpsummand}
-\frac{Gm_{j}m_{k}m_{p}(r_{j}-r_{k})\cdot(r_{j}-r_{p})}{\left\Vert r_{j}-r_{k}\right\Vert ^{3}}.
\end{equation}
Case 2: $\ell=k$ and $p\neq j,k$. This implies that $p>k$. Then $(r_{\ell}-r_{p})''=(r_{k}-r_{p})''$ where
\begin{equation}\label{eqkp}
(r_{k}-r_{p})''=\frac{-G(m_{k}+m_{p})(r_{k}-r_{p})}{\left\Vert r_{k}-r_{p}\right\Vert ^{3}}
+G\sum_{{i=1\atop i\neq k,p}}^{N}\frac{m_{i}(r_{i}-r_{k})}{\left\Vert r_{i}-r_{k}\right\Vert ^{3}}
-G\sum_{{i=1\atop i\neq k,p}}^{N}\frac{m_{i}(r_{i}-r_{p})}{\left\Vert r_{i}-r_{p}\right\Vert ^{3}}.
\end{equation}
Since $p\neq j,k$, only $i=j$ in second
summand on the right side of (\ref{eqkp}) gives rise to $r_{j}-r_{k}$, which
after taking the dot product with $m_{k}m_{p}(r_{k}-r_{p})$, leads to the summand 
\begin{equation}\label{kpsummand}
\frac{Gm_{j}m_{k}m_{p}(r_{k}-r_{p})\cdot(r_{j}-r_{k})}{\left\Vert r_{j}-r_{k}\right\Vert ^{3}}.
\end{equation}
Case 3: $\ell\neq k,j$ and $p=j$. This implies $\ell<j$. Then $(r_{\ell}-r_{p})''=(r_{\ell}-r_{j})''$ where
\begin{equation}\label{eqlj}
(r_{\ell}-r_{j})''=\frac{-G(m_{\ell}+m_{j})(r_{\ell}-r_{j})}{\left\Vert r_{\ell}-r_{j}\right\Vert ^{3}}
+G\sum_{{i=1\atop i\neq\ell,j}}^{N}\frac{m_{i}(r_{i}-r_{\ell})}{\left\Vert r_{i}-r_{\ell}\right\Vert ^{3}}
-G\sum_{{i=1\atop i\neq\ell,j}}^{N}\frac{m_{i}(r_{i}-r_{j})}{\left\Vert r_{i}-r_{j}\right\Vert ^{3}}.
\end{equation}
Since $\ell\neq j,k$, only $i=k$ in the
third summand on the right side of (\ref{eqlj}) gives rise to $r_{j}-r_{k}$,
which after taking the dot product with $m_{\ell}m_{j}(r_{\ell}-r_{j})$, leads to the summand
\begin{equation}\label{ljsummand}
\frac{Gm_{j}m_{\ell}m_{k}(r_{\ell}-r_{j})\cdot(r_{j}-r_{k})}{\left\Vert r_{j}-r_{k}\right\Vert ^{3}}.
\end{equation}
Case 4: $\ell\neq k,j$ and $p=k$. This implies $\ell<k$. Then $(r_{\ell}-r_{p})''=(r_{\ell}-r_{k})''$ where
\begin{equation}\label{eqlk}
(r_{\ell}-r_{k})''=\frac{-G(m_{\ell}+m_{k})(r_{\ell}-r_{k})}{\left\Vert r_{\ell}-r_{j}\right\Vert ^{3}}
+G\sum_{{i=1\atop i\neq\ell,k}}^{N}\frac{m_{i}(r_{i}-r_{\ell})}{\left\Vert r_{i}-r_{\ell}\right\Vert ^{3}}
-G\sum_{{i=1\atop i\neq\ell,k}}^{N}\frac{m_{i}(r_{i}-r_{k})}{\left\Vert r_{i}-r_{k}\right\Vert ^{3}}.
\end{equation}
Since $\ell\neq j,k$, only $i=j$ in the
third summand on the right side of (\ref{eqlj}) gives rise to $r_{j}-r_{k}$,
which after taking the dot product with the vector $m_{\ell}m_{k}(r_{\ell}-r_{k})$, lead to 
the summand
\begin{equation}\label{lksummand}
-\frac{Gm_{j}m_{\ell}m_{k}(r_{\ell}-r_{k})\cdot(r_{j}-r_{k})}{\left\Vert r_{j}-r_{k}\right\Vert ^{3}}.
\end{equation}
In all four cases, as evidenced by (\ref{jpsummand}), (\ref{kpsummand}),
(\ref{ljsummand}), and (\ref{lksummand}), there is a factor of the
form $m_{j}m_{k}m_{\alpha}$, where $\alpha$ is either $\ell$ or
$p$. We want to pairwise combine via the value of $\alpha$.
The above four cases imply that $\alpha\neq j,k$. However, $\alpha$
is free to be any other value from the set $\{1,...,N\}$. This leads
to following considerations: $\alpha<j$, $j<\alpha<k$, and $\alpha>k$.
Suppose $\alpha>k$. This occurs for all of the $p$ in Case 2 and
those $p$ in Case 1 for which $p>k$. We pairwise add (\ref{jpsummand})
and (\ref{kpsummand}) to obtain 
\small{
\begin{equation}\label{finalalpha>k}
-\frac{Gm_{j}m_{k}m_{p}(r_{j}-r_{p})\cdot(r_{j}-r_{k})}{\left\Vert r_{j}-r_{k}\right\Vert ^{3}} 
 +\frac{Gm_{j}m_{k}m_{p}(r_{k}-r_{p})\cdot(r_{j}-r_{k})}{\left\Vert r_{j}-r_{k}\right\Vert ^{3}}
  =-\frac{Gm_{j}m_{k}m_{p}}{\|r_{j}-r_{k}\|}.
\end{equation}}
\normalsize{Next consider $\alpha<j$.} This occurs for all of
the $\ell$ in Case 3 and those $\ell$ in Case 4 for which $\ell<j$.
We pairwise add (\ref{ljsummand}) and (\ref{lksummand}) to obtain
\small{
\begin{equation}\label{finalalpha < j}
\frac{Gm_{j}m_{\ell}m_{k}(r_{\ell}-r_{j})\cdot(r_{j}-r_{k})}{\left\Vert r_{j}-r_{k}\right\Vert ^{3}} 
 -\frac{Gm_{j}m_{\ell}m_{k}(r_{\ell}-r_{k})\cdot(r_{j}-r_{k})}{\left\Vert r_{j}-r_{k}\right\Vert ^{3}}
   =-\frac{Gm_{j}m_{k}m_{p}}{\left\Vert r_{j}-r_{k}\right\Vert },
\end{equation}}
\normalsize{where in the last term we relabeled $\ensuremath{\ell}$ to $\ensuremath{p}$.}
Finally we have to consider when $j<\alpha<k$. This occurs in the
remaining $p$ and $\ell$ of Cases 1 and 4 not covered by (\ref{finalalpha>k})
and (\ref{finalalpha < j}) respectively. We can pairwise add (\ref{jpsummand})
to (\ref{lksummand}), where in (\ref{lksummand}) we have renamed $\ensuremath{\ell}$ to $\ensuremath{p}$,
to obtain 
\small{
\begin{equation}\label{finalalphajkbetween}
-\frac{Gm_{j}m_{k}m_{p}(r_{j}-r_{p})\cdot(r_{j}-r_{k})}{\left\Vert r_{j}-r_{k}\right\Vert ^{3}} 
 -\frac{Gm_{j}m_{p}m_{k}(r_{p}-r_{k})\cdot(r_{j}-r_{k})}{\left\Vert r_{j}-r_{k}\right\Vert ^{3}}
 =-\frac{Gm_{j}m_{k}m_{p}}{\left\Vert r_{j}-r_{k}\right\Vert }.
\end{equation}}
\normalsize{If we add (\ref{finalalpha>k}) through (\ref{finalalphajkbetween})
together, we get $-G(M-m_j-m_k)m_jm_k/\|r_j-r_k\|$.
The above term is true for an arbitrary yet fixed $(j,k)$, where
$1\leq j<k\leq N$. By summing over $1\leq j<k\leq N$, we obtain
(\ref{simplifedT2jksum}).

\medskip
By combining (\ref{diffformkineticengy}) with (\ref{simplifedT2jksum})
we obtain 
\begin{equation}\label{desiredinequality}
\sum_{1\leq j<k\leq N}m_{j}m_{k}(r_{j}-r_{k})\cdot(r_{j}-r_{k})''
=-GM\sum_{1\leq j<k\leq N}\frac{m_{j}m_{k}}{\|r_{j}-r_{k}\|}<0.
\qquad \Box
\end{equation}

\section{The Restless Universe}
Below we list some consequences of Inequality (\ref{eq:NoConstantSolRelative}). These consequences reflect the fact
 that\\ $\sum_{1\leq j<k\leq N}m_{j}m_{k}(r_{j}-r_{k})\cdot(r_{j}-r_{k})''$ is always negative and that 
 for $1\leq j < k\leq N$ we made the implicit assumption
 of $r_j-r_k\not\equiv\overrightarrow{0}$ on $[a,b]$.  This last condition is tantamount to the collision free model.
\begin{corollary}\label{Inequalitycor}
\begin{enumerate}
\item{} Under the conditions of Theorem \ref{Inequalitythm}, the relative system (\ref{newtonseconddiff}) has no critical points
(constant solutions). 
\item{} For each $t\in[a,b]$ there exists an
index set $(j,k)$ with $1\leq j<k\leq N$ such that $(r_{j}(t)-r_{k}(t))''\neq\overrightarrow{0}$.
\item{}
For each $t\in[a,b]$ there exists an index set $(j,k)$ with
$1\leq j<k\leq N$ such that $r_{j}(t)-r_{k}(t)$ is not orthogonal to
$(r_{j}(t)-r_{k}(t))''$.
\item{} There exists an index set $(j,k)$ such that $(r_j-r_k)''\not\equiv\overrightarrow{0}$ on $[a,b]$.
Hence, there exists an index set $(j,k)$ such that $r_j-r_k$ is not identically constant on $[a,b]$; this implies
that one of the two position vectors associated with the index set $(j,k)$ is not identically constant on $[a,b]$.
\end{enumerate}
\end{corollary}

Corollary \ref{Inequalitycor}, Part 2, implies that 
$\left[
r''_1(t)-r''_2(t),
\hdots,
r''_i(t)-r''_j(t),
\hdots,
r''_{N-1}(t)-r''_N(t)
\right]^T \not\equiv \overrightarrow{0}$ for all $t\in[a,b]$.  However, as evidenced by the following corollary, we can say more.
\begin{corollary}\label{twopaircor}
 For $N\geq 3$, under the conditions of Theorem \ref{Inequalitythm}, 
 there are two distinct pairs of indices $(j_1, j_2)$ and $(j_3, j_4)$ from
 $\{(\ell,p)\}_{1\leq \ell < p \leq N}$
 such that  $(r_{j_1}-r_{j_2})''$ and $(r_{j_3}-r_{j_4})''$ are both not identically zero.
 \end{corollary}
{\bf Proof:} Without loss of generality, after relabeling if necessary, 
we can assume via (\ref{desiredinequality}) that $j_1 = 1$ and $j_2 = 2$ and that
$(r_1-r_2)''\not\equiv \overrightarrow{0}$.  Since
\[
\sum_{i=1}^{N-1}(r''_i-r''_{i-1}) + (r_{N}''-r_1'') =
(r_1''-r_2'') + (r_2''-r_3'') + \cdots + (r_{N-1}''-r_N) + (r_{N}''-r_1'') \equiv \overrightarrow{0},
\]
and since $(r_1-r_2)''\not\equiv \overrightarrow{0}$, then
\[
(r_2''-r_3'') + \cdots + (r_{N-1}''-r_N) + (r_{N}''-r_1'') \equiv - (r_1-r_2)''\not\equiv\overrightarrow{0},
\]
and we can choose $(j_3, j_4)$ to be any $(i, i+1)$ where $2\leq i\leq N-1$.\qquad $\Box$

\medskip
By combining Corollaries \ref{Inequalitycor} and \ref{twopaircor}, we deduce the following.
\begin{corollary}\label{twoaccelcor}
Let $N\geq 3$. Under the conditions of Theorem \ref{Inequalitythm}, for each $t\in[a,b]$, at least two bodies accelerate.
\end{corollary}

\begin{remark}
Assume that the $(r_j(t))_{j=1}^N$ associated with 
Theorem \ref{basicexistence theorem} are real analytic functions on $[c,d]$ where
$a\leq c < d\leq b$. 
 Furthermore assume that $r_j'' -r_k''\not\equiv \overrightarrow{0}$ on $[c,d]$.  
 Then $r_j''(t)-r_k''(t) =\overrightarrow{0}$ for only a finite number of $t\in[c,d]$.

\medskip
{\bf Proof By Contradiction:} Define $h(t):= r_j''(t)-r_k''(t)$ on $[c,d]$ and assume 
that $h(t_j) =\overrightarrow{0}$ for a countable number of $t_j\in[c,d]$.  There exists
a subsequence $(t_{j_\ell})$ of $(t_j)$ such that $t_{j_\ell}\rightarrow t_0\in[c,d]$ 
and  $h(t_{j_\ell}) = \overrightarrow{0}$.  The analyticity of $h$ at $t_0$ implies that
$h(t) = h(t_0) + \sum_{n=1}^{\infty}a_n(t-t_0)^n$
for some neighborhood of $|t-t_0|< \rho$ in the complex plane. By construction
$\mathrm{\lim}_{\ell\rightarrow\infty}h(t_{j_\ell}) = \overrightarrow{0}  = h(t_0)$.
Hence we deduce that
\begin{equation}\label{hpowerseries1}
h(t) = \sum_{n=1}^{\infty}a_n(t-t_0)^n = (t-t_0)\left[a_1 + \sum_{n=2}^{\infty}a_n(t-t_0)^{n-1}\right],\qquad |t-t_0| < \rho.
\end{equation}
If we substitute $t_{j_\ell}\neq t_0$ into (\ref{hpowerseries1}), since $t_{j_\ell}\neq t_0$ and $h(t_{j_\ell}) = \overrightarrow{0}$, 
we deduce that
\begin{equation}\label{powerseriesfactor}
 \overrightarrow{0} = a_1 + \sum_{n=2}^{\infty}a_n(t_{j_\ell}-t_0)^{n-1}
\end{equation}
Suppose $a_1\neq \overrightarrow{0}$. By analyticity of $h$ at $t_0$ there exists $\hat{\rho}$ such that 
\begin{equation}\label{lowerbound1}
0<\left|\sum_{n=2}^{\infty}a_n(t-t_0)^{n-1}\right|\leq \frac{|a_1|}{2},\qquad 0 < |t-t_0| < \hat{\rho} < \rho.
\end{equation}
The triangle inequality, when combined with (\ref{lowerbound1}) implies that
\begin{equation}\label{contradictionineq}
0 < \frac{|a_1|}{2} = |a_1| -  \frac{|a_1|}{2}\leq |a_1| - \left|\sum_{n=2}^{\infty}a_n(t-t_0)^{n-1}\right|
\leq \left|a_1 + \sum_{n=2}^{\infty}a_n(t-t_0)^{n-1}\right|,
\end{equation}
whenever $0 < |t-t_0| < \hat{\rho} < \rho$.  Since $t_{j_\ell}\rightarrow t_0$, Inequality (\ref{contradictionineq}) holds for
those $t_{j_\ell}\in |t-t_0| < \hat{\rho}$ and implies
that
$a_1 + \sum_{n=2}^{\infty}a_n(t_{j_\ell}-t_0)^{n-1}\not\equiv\overrightarrow{0}$,
a contradiction of (\ref{powerseriesfactor}).  Hence $a_1 = \overrightarrow{0}$.
By repeating this argument, we show 
there is a neighborhood of $t_0$ in the complex plane such that all $a_n\equiv \overrightarrow 0$, i.e
$h(t)\equiv \overrightarrow 0$ for $|t-t_0|< \rho_1$.  
Then by analytic continuation we conclude that $h(t)\equiv \overrightarrow{0}$ for $t\in[c,d]$,
a contradiction of the initial assumption. \qquad $\Box$
\end{remark}

\section{Conclusions}
This article is motivated by the necessity to have a framework
of equations for celestial mechanics which support measurements from any
point mass in the universe whether accelerating or not. In Section 2 we showed that  
 NCME for the two and three body problems do not satisfy this requirements.
 In order to construct a framework of origin invariant equations, we derive from NCME a relative
system of equations, either System $(RS1)$ or System $(RS2)$, that is consistent in any CCS. System $(RS1)$ 
(respectively System $(RS2)$) is shown
to be WPOACS, to accommodate BCOS, and be invariant with respect to
an arbitrary time dependent translation of coordinates of the form
$r_{i}(t)=\widetilde{r}_{i}(t)-c(t),\:c''(t)\neq\overrightarrow{0},\:t\in[a,b]$; see Theorem
\ref{origininvariantNbodythm}.
We recall that NCME are invariant only with respect to inertial translations
where $c''(t)\equiv\overrightarrow{0}$. NCME and $(RS1)$
are related but are not equivalent. However, for the earth-sun Kepler problem,
the NCME and $(RS1)$ are approximately the same. 
The adoption of the relative system comes with an uncertainty advisory. 
Given an $N$-body problem we can be certain about the relative position of $N-1$-point masses
 but we must stay uncertain about the position of one point mass.  
Finally
we observe that we need a substitute for the resultant of forces acting
on the binary mass $(m_{j},m_{k})$ associated with the relative accelerations
$(r_{j}-r_{k})''$ of $(RS1)$ (respectively $(RS2)$). The integral
of motion calculations in \cite{relmotion} suggests the use of a
factor $\mu m_{j}m_{k}$, where $\mu$ is a certain cosmological constant
to be determined by experiment.

\medskip
We end this article with a foundational proposition which underlies the calculations of Section 2.
\begin{proposition}\label{foundationprop}
Given a unique CCS whose origin $O$ corresponds to a point mass $m$, if $r(t)\in\mathbb{R}^3$ is defined to be
the position of $m$ as measured from $O$, and if $A(t)\in C^2(\mathbb{R})$, then the initial value problem
\begin{equation}\label{eq:MODELMECHANICSPROBLEM}
r''(t)=A(t),\qquad r(t_{0}),\,r'(t_{0})\in\mathbb{R}^3
\end{equation}
does not have a solution unless $A(t)\equiv \overrightarrow{0}$.
\end{proposition}
The proof follows from the fact the assumptions imply that $r(t)\equiv r'(t)\equiv r''(t) \equiv \overrightarrow{0}$.
The inconsistencies shown in Section 2 are a result of the assumptions of Proposition \ref{foundationprop}.
A main result of this article is employing relative systems as a means of avoiding the contradiction inherent in Proposition \ref{foundationprop}.
These relative difference systems utilize only one CCS with an origin O that coincides with a point
mass $m_{j}$ and are shown to be WPOACS.   If $A(t)\not\equiv\overrightarrow{0}$, then a solution to (\ref{eq:MODELMECHANICSPROBLEM})
exists if and only if the assumptions of Proposition \ref{foundationprop} are invalid.  The negation of these assumptions logically corresponds to 
the following statement:
{\it either there is a unique CCS whose origin $O$ does not correspond to a point mass {\bf or} there is more than one CCS.}
This leads to the following corollary.
\begin{corollary}\label{foundationpropneg}
Given a unique CCS whose origin $O$ does not correspond to any point mass $m$, if $r(t)\in\mathbb{R}^3$ 
is defined to be
the position of $m$ as measured from $O$, and if $A(t)\in C^2(\mathbb{R})$, then the initial value problem
\begin{equation}\label{eq:MODELMECHANICSPROBLEM1}
r''(t)=A(t)\not\equiv\overrightarrow{0},\qquad r(t_{0}),\,r'(t_{0})\in\mathbb{R}^3
\end{equation}
has a unique solution given by $r(t) = r(t_0) + r'(t_0)(t-t_0)+\int_{t_0}^t\int_{t_0}^sA(u)\,du\,ds$.
\end{corollary}
\begin{corollary}\label{foundationpropnega}
Given two distinct CCSs, one with origin $O$ and the other with origin $\widetilde{O}\neq O$, set 
$r(t):=\overrightarrow{O\widetilde{O}}(t)$.  Then the initial value problem (\ref{eq:MODELMECHANICSPROBLEM1})
has a unique solution.
\end{corollary}
The existence of more than one CCS provides 
another means, independent of the relative difference WPOACS,
 in which to have the origin of a coordinate system coincide with a point mass in a manner which avoids
 the contradictions of Proposition \ref{foundationprop}.
 Assume we want our equations
formulated in a CCS whose origin $\widetilde{O}$ coincides with the point
mass $m_{1}$.  We start with NCME (\ref{newtonlawcm}) that are formulated
in a CCS with origin $O$ that does not coincide with any of the $N$
point masses. Theorem \ref{basicexistence theorem} guarantees
a solution on some interval $[a,b]$. 
If we let $\tilde{r}_i(t)$ denote the position of $m_i$ with respect to $\widetilde{O}$,
since $r_i(t) = r_1(t) + \tilde{r}_i(t)$, we transform (\ref{newtoninitalvaluecm}) into
\begin{equation}\label{eq:r1tilda}
r''_{1}(t)+\widetilde{r}''_{1}(t)=\sum_{j=2}^N\frac{Gm_{j}(r_{j}(t)-r_{1}(t))}{\|r_{j}(t)-r_{1}(t)\|^{3}}
\equiv\sum_{j=2}^N\frac{Gm_{j}\widetilde{r}_{j}(t)}{\|\widetilde{r}_{j}(t)\|^{3}},
\end{equation}
and
\begin{equation}\label{eq:ritilda}
r''_{1}(t)+\widetilde{r}''_{i}(t)=\sum_{j\neq i}\frac{Gm_{j}(r_{j}(t)-r_{i}(t))}{\|r_{j}(t)-r_{i}(t)\|^{3}}
\equiv\sum_{j\neq i}\frac{Gm_{j}(\widetilde{r}_{j}(t)-\widetilde{r}_{i}(t))}{\|\widetilde{r}_{j}(t)-\widetilde{r}_{i}(t)\|^{3}}
,\;2\leq i\leq N,
\end{equation}
which is a different set of equations than the relative equations in
of Section 3. Since\\ $r''_1(t) = \sum_{j=2}^N\frac{Gm_{j}(r_{j}(t)-r_{1}(t))}{\|r_{j}(t)-r_{1}(t)\|^{3}}$, 
Equation (\ref{eq:r1tilda}) is equivalent to $\tilde{r}_1(t) = 0$.
The system of equations provided by (\ref{eq:r1tilda})
is equivalent to postulating that
 the $N$-body problem is governed by a system of
equations 
\[
r''_{i}(t)+c''(t)=\sum_{j=1\atop j\neq i}^N\frac{Gm_{j}(r_{j}(t)-r_{i}(t))}{\|r_{j}(t)-r_{i}(t)\|^{3}},\;1\leq i\leq N,
\]
where $c(t)\in C^{2}([a,b])$ is a fictitious acceleration that is yet to be determined, and where
$[a,b]$ is an interval of existence of the desired solutions. This
will allow for non inertial translations, but in a manner different than that provided by
Theorem \ref{origininvariantNbodythm}.


\begin{thebibliography}{99}
\bibitem{Bannikova} E. Bannikova and M. Capaccioli, \textit{Foundations
of Celestial Mechanics}, Graduate Texts in Physics, 1st Edition, Springer,
2022. 

\bibitem{HSIEHSIBUYA} P. F. Hsieh and Y. Sibuya, \textit{Basic Theory
of Ordinary Differential Equations}, Universitext, Springer-Verlag,
New York, 1999.

\bibitem{Pollard} H. Pollard, \textit{Celestial Mechanics}, Carus
Mathematical Monographs, No. 18, Mathematical Association of America,
1976.

\bibitem{relmotion} J. Quaintance and H. Gingold, \textit{Relative
Difference System Involving Newton's Equations of Celestial Mechanics},
Preprints 2022, 2022100277. 

\bibitem{Wintner} A. Wintner, \textit{The Analytical Foundation of
Celestial Mechanics}, Princeton University Press, 1941. 
\end{thebibliography}
\end{document}